\documentclass[pra,amssymb,amsfonts,twocolumn]{revtex4}
\usepackage{epsfig}
\usepackage{bm}
\usepackage{amsmath}
\usepackage{amsfonts}
\usepackage{amssymb}

\def\beq{\begin{equation}}
\def\eeq{\end{equation}}
\def\bea{\begin{eqnarray}}
\def\eea{\end{eqnarray}}
\def\half{\mbox{$1\over2$}}

\def\halft{\mbox{$\theta\over2$}}
\def\qtr{\mbox{$1\over4$}}
\def\ses{\mbox{$1\over16$}}

\def\bK{{\bf K}}
\def\bL{{\bf L}}
\def\bJ{{\bf J}}

\def\bk{{\bf k}}
\def\bu{{\bf u}}
\def\bv{{\bf v}}
\def\bz{{\bf z}}
\def\bn{{\bf n}}
\def\bx{{\bf x}}
\def\by{{\bf y}}
\def\bw{{\bf w}}
\def\br{{\bf r}}
\def\bb{{\bf b}}
\def\cD{{\cal D}}

\def\hbb{{\hat{\bb}}}
\def\hbr{{\hat{\br}}}
\def\hbx{{\hat{\bx}}}
\def\hby{{\hat{\by}}}

\def\tal{{\tilde\theta}}
\def\teb{{\tilde\phi}}
\def\tap{{\tilde\varphi}}

\def\bl{\mbox{\boldmath $\ell$}}
\def\1{\mbox{1\hskip-.25em l}}
\def\6{\langle }
\def\9{\rangle }

\begin{document}
%\draft

\title{Elliptic Rydberg States as Direction Indicators}
\author{Netanel H. Lindner, Asher Peres, and  Daniel R. Terno%
\footnote{Now at Perimeter Institute, Waterloo, Ontario, Canada}}
\address{Department of Physics, Technion---Israel Institute of
Technology, 32 000 Haifa, Israel}
%\maketitle
\begin{abstract}
The orientation in space of a Cartesian coordinate system can be
indicated by the two vectorial constants of motion of a classical
Keplerian orbit: the angular momentum and the Laplace-Runge-Lenz
vector. In quantum mechanics, the states of a hydrogen atom that
mimic classical elliptic orbits are the coherent states of the
SO(4) rotation group. It is known how to produce these states
experimentally. They have minimal dispersions of the two conserved
vectors and can be used as direction indicators. We compare the
fidelity of this transmission method with that of the idealized
optimal method.
\bigskip
\end{abstract}
\pacs{PACS numbers: 03.67.Hk, 03.65.Ta, 03.65.Ud}
\maketitle
\section{Unspeakable quantum information}

Information theory usually deals with the transmission of a
sequence of discrete symbols, such as 0 and 1. Even if the
information to be transmitted is of continuous nature, such as the
position of a particle, it can be represented with arbitrary
accuracy by a string of bits. However, there are situations where
information cannot be encoded in such a way. For example, the
emitter (conventionally called Alice) wants to indicate to the
receiver (Bob) a direction in space. If they have a common
coordinate system to which they can refer, or if they can create
one by observing distant fixed stars, Alice simply communicates to
Bob the components of a unit vector \bn\ along that direction, or
its spherical coordinates $\theta$ and $\phi$. But if no common
coordinate system has been established, all she can do is to send
a real physical object, such as a gyroscope, whose orientation is
deemed stable.

In the quantum world, the role of the gyroscope is played by a
system with large spin. For example, Alice can send angular
momentum eigenstates satisfying $\bn\cdot\bJ|\psi\9=j|\psi\9$.
This is essentially the solution proposed by Massar and
Popescu~\cite{mp} who took $N$ parallel spins, polarized
along~\bn. This, however, is not the most efficient procedure: for
two spins, a higher accuracy is achieved by preparing them with
opposite polarizations \cite{gp}. For more than two spins, optimal
results are obtained with entangled states
\cite{peres1,catalans1}.

The above discussion can be generalized to the transmission of a
Cartesian frame. If $N$ spins are available, one can encode a
Cartesian frame in an entangled state of these spins, as in
\cite{catalans2}. However, a more accurate transmission is then
obtained if Alice uses half of the spins to indicate the $x$-axis,
and the other half for her $y$-axis \cite{cov}. In this case the
$x$ and $y$ directions found by Bob may not be exactly
perpendicular and some adjustment will be needed to obtain Bob's
best estimate of the $x$ and $y$ axes. Finally, the $z$-axis can
be inferred from the estimates of the $x$ and $y$ axes.

However, it is not possible to proceed in this way if a {\it
single\/} quantum messenger is available. The optimal transmission
of a Cartesian frame by a hydrogen atom (formally, a spinless
particle in a Coloumb potential) was derived by Peres and Scudo
\cite{peres2}. The results of
\cite{catalans2} can also be used for a hydrogen atom if one considers
the angular momentum eigenstates to be those of the atom, rather
than those of $N$ spins.

In this paper we show how to transmit a Cartesian frame by using
the elliptic Rydberg states of a hydrogen atom. These are the
quantum mechanical analog of a classical Keplerian orbit, and it
is known how to produce these states experimentally.  Elliptic
Rydberg states, just as their classical counterparts, define three
orthogonal directions in space, and thus are natural candidates
for encoding a Cartesian frame.

In the following section we discuss the properties of quantum
elliptic states. Section III deals with the transmission of one
direction by means of them, and in Sec.~IV we use them to transmit
two orthogonal axes (and thus a Cartesian frame). In Sec.\ III and
IV the detection procedure is based on SO(3) coherent states as in
\cite{peres2}. SO(4) coherent states are employed in Sec.~V to
produce a positive operator valued measure (POVM) which enables
the use of elliptic states for the transmission of two directions
that are not orthogonal. Even when these states are used by Alice
to transmit two orthogonal axes, the two directions found by Bob
are not necessarily perpendicular and further adjustment is
needed, as explained above. As shown in the Appendix, these
adjustments increase the fidelity of transmission of two
orthogonal axes. However, a higher fidelity is achieved with a
POVM based on the SO(3) rotation group, especially when the energy
quantum number $n$ is large.

\section{Construction of an elliptic state}
A classical bounded Keplerian orbit in a potential $-k/r$  can be
defined by its constants of motion: the energy $E<0$, the angular
momentum $\bL$ which is an axial vector perpendicular to the plane
of the orbit, and the Laplace-Runge-Lenz (LRL) vector \cite{gold}
\beq
\bK = (-2H)^{-1/2}\left({\bf p}\times{\bf L}-\mu k{\bf r}/r\right),
\eeq
where $\mu$ is the particle's reduced mass, and we introduced a
prefactor $(-2H)^{-1/2}$ for later convenience. This prefactor,
which is a constant of motion, does not appear in the usual
definition of the LRL vector
\cite{gold}. The Hamiltonian is $H={\bf p}^2/2m -k/r$.
We consider only bounded motion for which the energy $E$, which is
the numerical value of $H$, is negative. The classical orbit is
then an ellipse, and the LRL vector is a polar vector directed
along its major axis. It satisfies:
\beq
{\bf L}\cdot{\bf K} = 0,
\label{classical1}
\eeq
and
\beq
{\bf K}^2 + \mu {\bf L}^2 = -\mu^2k^2 /2E.
\label{classical2}
\eeq
Because of these relations, only five out of the seven constants
are independent. They uniquely determine the shape and orientation
of the ellipse. Its eccentricity is \cite{gold}
\beq
e =|\bK| \sqrt{-2E} /\mu k.
\label{eccentricity}
\eeq

We now turn to the quantum version. We use natural units, $\mu = k
=\hbar = 1$, so that the energy levels for bound states are
$E=-1/2n^2$, where $n$ is an integer. The operator $\bK$ is
defined by

\beq
{\bf K} = (-2H)^{-1/2}\left(\half({\bf p}\times{\bf L}-{\bf
L}\times {\bf p})-{\bf r}/r\right).
\eeq

\noindent Note that $H$ commutes with $\left(\half({\bf p}\times{\bf
L}-{\bf L}\times {\bf p})-{\bf r}/r\right)$. The commutation
relations for the operators ${\bf L}$ and ${\bf K}$ are
\cite{landau}
\beq
[L_{i},K_{j}]=i \epsilon_{ijk}K_{k},
\label{com1}
\eeq
\beq
[K_{i},K_{j}]=i \epsilon_{ijk}L_{k}.
\label{com2}
\eeq

Together with $[L_{i},L_{j}]=i \epsilon_{ijk}L_{k}$, these are the
commutation rules of infinitesimal rotations in four dimensional
Euclidean space, which leave the $n$-th energy level subspace
invariant \cite{rmp}. The coherent states of SO(4), {\it i.e.} the
states for which the dispersion of ${\bf L}^2 + {\bf K}^2$ is
minimal, can be built from the coherent states of SO(3), since
SO$(4)=\,$SO(3)$\times$SO$(3)$.

Define
\beq
{\bf J}_1 = \half({\bf L} - {\bf K})\qquad{\rm and}\qquad{\bf
J}_2=\half ({\bf L} + {\bf K}).
\label{j2}
\eeq
These two operators have the commutation relations of two
independent three-dimensional angular momenta:
\beq
[J_{1i}, J_{1j}]=i \epsilon_{ijk}J_{1k},
\eeq
\beq
[J_{2i}, J_{2j}]=i \epsilon_{ijk}J_{2k},
\eeq
\beq
[J_{1i}, J_{2j}]=0.
\eeq
Instead of the classical equations (\ref{classical1}) and
(\ref{classical2}) we now have \cite{delande}
\beq
{\bf L}\cdot{\bf K} = {\bf K}\cdot{\bf L} = 0,
\eeq
and
\beq
{\bf L}^2 + {\bf K}^2 = -1 -1/2H = n^2-1,
\label{quantum2}
\eeq
where the last form of the equality holds for energy eigenstates.
In the classical limit, $n\gg 1$, Eq.~(\ref{quantum2}) reduces to
the classical one~(\ref{classical2}).

In the rest of this article we consider only energy eigenstates.
Owing to Eq.~(\ref{j2}), we have
\beq
j_1(j_1+1) = j_2(j_2+1)= \qtr(n^2-1).
\label{j1=j2}
\eeq
where $j_1$ and $j_2$ are the quantum numbers referring to the
operators $\bJ_1^2$ and $\bJ_2^2$ respectively. It follows that
$j_1$ and $j_2$ are equal: $j_1=j_2\equiv j$, and $j(j+1)=
\qtr(n^2-1)$, so that
\beq
j=\half(n-1).
\eeq
 The coherent
states of a three-dimensional angular momentum will be denoted by
$|{J,\textbf{u}\9}$. They obey $\textbf{u}\cdot\textbf{J}|{J,
\textbf{u}\9}=j|{J,\textbf{u}\9}$ for an arbitrary classical unit vector
$\bu$.
For the coherent states the dispersion $\Delta{\bf J} =
(\6\textbf{J}^2\9 -
\6\textbf{J}\9^2)^{1/2}$ is minimal: $(\Delta{\bf J})^2
= j$. In particular, $\Delta{J_u} = 0$, and $\Delta{J_{\bot}} =
\sqrt{j/2}$, where $J_u={\bf J}\cdot\bu$ and
$J_{\bot}={\bf J}\cdot\bv$ with $\bv\perp\bu$. The coherent states
of SO(3) are obtained by a rotation of a fiducial coherent state
$|{J,\textbf{z}\9}$,
\beq
|{J, \bu_{\theta
\phi}\9} = e^{-i L_z \phi} e^{-i L_y \theta}|{J, \textbf{z}\9}.
\eeq

The coherent states of SO(4) are now obtained as direct products
of coherent states for each of the SO(3) subgroups,
\beq
|{n, \textbf{u}_1 \textbf{u}_2\9} = |{J_1, \textbf{u}_1\9} \otimes
|{J_2, \textbf{u}_2\9}.
\label{product}
\eeq
where the unit vectors $\bu_1$ and $\bu_2$ are again classical.
The coherent state $|{J_1, \bu_1\9}$ obeys
\beq
\bu_1\cdot\bJ_1|{J_1, \bu_1\9}=j|{J_1, \bu_1\9}=\half(n-1)|{J_1, \bu_1\9},
\label{defu1}
\eeq
and likewise we have
\beq
\bu_2\cdot\bJ_2|{J_2, \bu_2\9}=j|{J_2, \bu_2\9}=\half(n-1)|{J_2,
\bu_2\9}.
\label{defu2}
\eeq
As from now, we shall omit the symbols $n, J_1,$ and $J_2$ in
state vectors, since the quantum numbers $n,j_1,j_2$ have fixed
values, related by Eq.~(\ref{j1=j2}). For example the state
$|{nlm\9}$ which obeys $H|{nlm\9}=n|{nlm\9}$,
$\bL^2|{nlm\9}=l(l+1)|{nlm\9}$, and $L_z|{nlm\9}=m|{nlm\9}$ will
be written simply as $|{lm\9}$, $|{J,
\textbf{u}\9}$ becomes $|{\textbf{u}\9},$ and $|{J,
\textbf{z}\9}$ becomes $|{jj\9}$, etc. The
symbol $j$ will always denote the fixed value $j=\half(n-1)$.

Owing to Eq.~(\ref{j2}), the dispersion of ${\bf L}^2 + {\bf K}^2$
is minimal for coherent states \cite{delande}:
\beq
(\Delta{\bf L})^2 + (\Delta{\bf K})^2 = 2[(\Delta{\bf J}_1)^2 +
(\Delta{\bf J}_2)^2] =2(n-1).
\eeq
To obtain the expansion of the coherent state $|{n,
\textbf{u}_1 \textbf{u}_2\9}$ in the familiar $nlm$ basis, we first
expand each of the $|{\textbf{u}_i\9}$ in Eq.~(\ref{product}):
\beq
|\bu_i\9=\sum_{m=-j}^{j} D^{j}_{m} (\theta_i\,\phi_i) |{j m \9},
\qquad i=1,2,
\label{ji}
\eeq
where the $ D^{j}_{m} (\theta_i\,\phi_i)$ are related to the usual
rotation matrices \cite{edmonds}:
\bea
\!\!\!\!\!\!\!\!\!D^{j}_{m} (\theta \phi)&\!\!\equiv&\!\cD^{\,(j)}(\phi
\theta 0)_{m\,j},\\
 \hspace*{-5mm} &\!\!\!=&
\!\!\!\!{{2j}\choose{j+m}}^{1/2}\!\!(\cos\halft)^{j+m}
 (\sin\halft)^{j-m} \,e^{-im\phi}.  \eea
Substitution into Eq.~(\ref{product}) gives
\beq
|{\textbf{u}_1 \textbf{u}_2\9}
=\sum_{m_1=-j}^{j}\;\sum_{m_2=-j}^{j} D^{j}_{m_1}(\theta_1
\phi_1) D^{j}_{m_2}(\theta_2
\phi_2)|{j m_1\9}\otimes |{j m_2\9}.
\label{almosthere}
\eeq
We then use the angular momentum addition formula,
\beq
|{j m_1\9}\otimes |{j m_2\9} =
\sum_{l=0}^{2j}\sum_{m=-l}^{l} C^{jj\,l}_{m_1m_2\,m} |{lm\9},
\label{clebsch}
\eeq
where $C^{j_1j_2\,l}_{m_1m_2\,m}$ is the Clebsch-Gordan
coefficient
\cite{tinkham} which vanishes for $m\not= m_1+m_2$. Combining
Eqs.~(\ref{almosthere}) and~(\ref{clebsch}),  we finally get
\begin{widetext}
\beq
|{\textbf{u}_1 \textbf{u}_2\9}=\sum_{l=0}^{2j}\sum_{m=-l}^{l}\;
\left(\sum_{m_1=-j}^{j}\;\sum_{m_2=-j}^{j} D^{j}_{m_1}(\theta_1
\phi_1) D^{j}_{m_2}(\theta_2
\phi_2)C^{jj\,l}_{m_1m_2\,m}\right)|{lm\9}.
\eeq
\end{widetext}
The classical orbit that corresponds to the coherent state
$|{\textbf{u}_1 \textbf{u}_2 \9}$, in the limit of large $n$, can
be obtained as follows. From (\ref{j2}), we have
\beq
{\bf L} = ({\bf J_1} + {\bf J_2})\qquad{\rm and}\qquad  {\bf K}=
({\bf J_2} - {\bf J_1}).
\label{LA}
\eeq
Let $\zeta$ be half the angle between $\bu_1$ and $\bu_2$, {\it
i.e.}
 $\sin\zeta\equiv|\textbf{u}_1\times \textbf{u}_2| /
|\textbf{u}_1+
\textbf{u}_2|$, and define three orthogonal classical unit vectors
\beq
\bl\equiv\frac{\textbf{u}_1 + \textbf{u}_2}{|\textbf{u}_1 +
\textbf{u}_2|},\qquad \textbf{k} \equiv
\frac{\textbf{u}_2 - \textbf{u}_1}{|\textbf{u}_2 -
\textbf{u}_1|},
\label{kl}
\eeq
and $\textbf{w} \equiv \bl \times
\textbf{k}$. Denoting by
$\bu_{1\perp}$ an arbitrary vector orthogonal to $\bu_1$, we have
\beq
\bJ_1 = (\bJ_1\cdot\bu_1)\bu_1+(\bJ_1\cdot\bu_{1 \perp})\bu_{1
\perp},
\eeq
\beq
\bJ_1\cdot\bu_2
=(\bJ_1\cdot\bu_1)(\bu_1\cdot\bu_2)+(\bJ_1\cdot\bu_{1
\perp})(\bu_{1 \perp}\cdot\bu_2).
\eeq
Then from
\beq
{\6\bu_1|}(\bu_1\cdot\bJ_1)|{\bu_1\9} = j ,\eeq and
\beq {\6\bu_1|}\bu_{1\perp}\cdot\bJ_1|{\bu_1\9}=0, \eeq
we get
\bea {\6 \bu_1|}(\bu_2\cdot\bJ_1)|{\bu_1\9}&=&{\6
\bu_2}|(\bu_1\cdot\bJ_2)|{\bu_2\9},\\
 &=&j\,\bu_1\cdot\bu_2=j\cos2\zeta.  \label{melement} \eea
Noting that
\beq |\bu_1+\bu_2| = 2|\cos\zeta|,\eeq
and
\beq |\bu_1-\bu_2| = 2|\sin\zeta|, \label{length} \eeq
we obtain from (\ref{melement}) the expectation values of the
components of ${\bf K}$ and ${\bf L}$ along the directions of
$\bk$, $\bl$ and $\bw$, for the coherent state $|\bu_1 \bu_2\9$
\beq
\6K_k \9 = (n-1) \sin\zeta,
\label{expK}
\eeq
\beq
\6L_\ell \9 = (n-1)\cos\zeta,
\label{expL}
\eeq
where $K_k\equiv\bk\cdot\bK$, etc. In the perpendicular directions
the expectation values vanish:
\beq
\6K_\ell\9 = \6K_w\9 = \6L_k\9 = \6L_w\9 = 0.
\label{experp}
\eeq

\noindent From Eqs.~(\ref{expK})--(\ref{experp}) we see that in the limit
of large
$n$, the coherent state $|{\textbf{u}_1 \textbf{u}_2 \9}$
corresponds to a classical elliptic trajectory in the $\bk\bw$
plane with the LRL vector in the {\bf k} direction, the angular
momentum in the $\bl$ direction, and eccentricity
$e=\6K_k\9/(n-1)=
\sin\zeta$, which is the quantum mechanical analog to
Eq.~(\ref{eccentricity}). The unit vector $\bw$ is parallel to the
minor axis of the ellipse.

\section{ Transmission of one direction}
We now turn to the use of elliptic wave functions as direction
indicators. Consider two observers (Alice and Bob) who do not have
a common reference frame. Alice wants to indicate to Bob her
$z$-axis by using a hydrogen atom in a Rydberg state. In the next
section, we shall likewise discuss the transmission of two
orthogonal axes by a single hydrogen atom. We use as much as
possible the same notations in both sections. Alice's signal is
\beq
|{A\9} = \sum_{l=0}^{2j}\;\sum_{m=-l}^{l} a_{lm} |{l m\9},
\label{alicegeneral}
\eeq
where
\beq
\sum_{l=0}^{2j}\;\sum_{m=-l}^{l} |a_{lm}|^2 = 1.
\eeq

Bob's detectors have labels $\psi \theta \phi$ which indicate the
unknown Euler angles relating his Cartesian axes to those used by
Alice. The mathematical representation of his apparatus is a POVM
\beq \int dE(\psi\theta\phi)=\1, \eeq
where
\beq
 dE(\psi\theta\phi)=d_{\psi\theta\phi}
 U(\psi\theta\phi)|B\9\6B|U^{\dag}(\psi\theta\phi),\label{povm}
\eeq
and $d_{\psi \theta \phi} = \sin\theta d\psi d\theta d\phi /8
\pi^2$ is the SO(3) Haar measure for Euler angles
\cite{edmonds}. As usual,
$U(\psi \theta \phi)$ is the unitary operator for a rotation by
Euler angles $\psi \theta \phi$, and $|{ B\9}$ is Bob's fiducial
vector defined as in \cite{peres2},
\beq
\,|{ B\9} = \sum_{l=0}^{2j}\sqrt{2l+1}\sum_{m=-l}^{l}
b_{lm}\,|{lm\9},
\label{bobgeneral}
\eeq
\noindent where for each $l$,
\beq
\sum_{m=-l}^{l} |b_{lm}|^2 =1.
\eeq
Note that Eq.~(\ref{alicegeneral}) was written with Alice's
notation, while Eq.~(\ref{bobgeneral}) is in Bob's notation
(recall that they use different coordinate systems).

Optimizing the transmission fidelity, defined by
Eq.~(\ref{fidelity}) below, leads to
\cite{peres2}
\beq
b_{lm} = a_{lm}\left(\sum_{n=-l}^{l}|a_{ln}|^2\right)^{-1/2},
\label{bob}
\eeq
for each $l$. Since $|{ B\9}$ is a direct sum of vectors, one for
each value of $l$, then likewise $U(\psi
\theta
\phi)$ is a direct sum with one term for each irreducible
representation:
\beq
U(\psi \theta \phi) =\sum_{l}\oplus \cD^{\,(l)}(\psi \theta \phi),
\eeq
where the $\cD^{\,(l)}(\psi \theta \phi)$ are the usual
irreducible unitary rotation matrices \cite{edmonds}. A
generalization of Schur's lemma \cite{wigner} confirms that
Eq.~(\ref{povm}) is indeed satisfied, owing to the coefficients
$\sqrt{2l+1}$ in Eq.~(\ref{bobgeneral}).

The fidelity of the transmission of a single direction is defined
as usual:
\beq
F = \6\cos^2(\omega/2)\9 = \half (1+ \6\cos\omega\9),
\label{fidelity}
\eeq
where $\omega$ is the angle between the direction indicated by
Alice and the one that is estimated by Bob. If Alice indicates her
$z$-axis, we thus want to maximize $\6\cos\omega_z\9$. Following
the method of Peres and Scudo
\cite{peres2} we define Euler angles $\alpha
\beta \gamma$ whose effect is rotating Bob's Cartesian frame
into his {\it estimate} of Alice's frame, and then rotating back
the result by the {\it true} angles from Alice's to Bob's frame.
The angles $\alpha
\beta \gamma$ thus indicate Bob's measurement error.
Since in this case Bob's estimate refers to Alice's $z$-axis only,
the angle $\omega_z$ is identical to the second Euler angle
$\beta$:

\beq
\6 \cos\omega_z\9=\int d_{\alpha
\beta \gamma}|{\6 A}|U(\alpha
\beta \gamma)|{ B\9}|^2\cos\beta.
\label{expectation}
\eeq

Let us  examine two extreme cases. First, we take Alice's vector
to be a circular state, with null eccentricity ($\sin{\zeta}=0),$
{\it i.e.},
\beq
|{A \9}= |{ ll\9},
\eeq
with
\beq
l = 2j= n-1.
\eeq
Bob's vector is obtained from (\ref{bob}), which in this case
gives
\beq
|{ B\9}=\sqrt{2n-1}\,\,|{l l\9}.
\eeq

We then have \cite{edmonds}
\bea
\!\!\!{\6 A}|U(\alpha\beta \gamma)|{ B\9} =
\sqrt{2n-1}\,e^{i(n-1)(\alpha+\gamma)} \cos^{\,2(n-1)}{\beta\over2}.
\eea
Inserting the last equation into Eq. (\ref{expectation}) gives
\bea
\!\!\!\!\!\!\!\!\!\6\cos\omega_z\9&\!\!=&\!\!(n-\half)\int_{0}^{\pi}
\sin\beta d\beta
\cos^{\,4(n-1)}(\beta/2) \cos\beta, \label{wzcircular}
\\&\!\!=&\!\!(n-1)/n.
\eea
The ``infidelity" $1-F$, whose typical meaning is Bob's mean
square error \cite{peres1}, is
\beq \half(1-\6\cos\omega_z\9)=1/2n.  \label{msecirc} \eeq
This result is identical to the one obtained by Massar and Popescu
\cite{mp}, when the number of their parallel spins is taken to be
$N=2l=2n-2$.

The other extreme case corresponds to a classical elliptical orbit
with unit eccentricity, so that $\bL=0$. Let $\bK$ lie in the
$\bz$ direction. The corresponding quantum state, denoted by
$|{K,\bz\9}$, is an extreme Stark state with $\6L_z\9=0$ and
maximal ${\6 K_z\9}$:
\beq
|{K, \textbf{z}\9} \equiv |{-\textbf{z}\9}\otimes |{\textbf{z}\9}
= |{j,-j\9} \otimes|{j j\9}.  \eeq This state satisfies
$L_z|{K,\textbf{z}\9}=0$ and it is also an eigenstate of $K_z$:
\begin{eqnarray}
K_z|{K, \textbf{z}\9}&=&(J_{2z}-J_{1z})|{j,-j\9}
\otimes|{j j\9},\\ &=& (n-1)|{K, \textbf{z}\9}, \end{eqnarray}
owing to
\beq(J_{2z}-J_{1z})|{j,{-j}\9} \otimes|{j j\9}=-J_{1z}|{j,-j\9}
\otimes J_{2z}|{j j\9}. \eeq
In the $nlm$ basis we have
\beq |{K, \textbf{z}\9}
=\sum_{l=0}^{2j}\sum_{m=-l}^{l}C^{jj\,l}_{m_1m_2m}|{lm\9}
=\sum_{l=0}^{2j}C^{jj\,l}_{-j j 0}|{l0\9}, \label{extreme} \eeq
since $m_1 = -j$ and $m
_2 = j$. The fidelity of transmission by
this state will be evaluated at the end of this Section.

Both the circular state and the extreme Stark state are coherent
states of SO(4), but only the circular state is also an angular
momentum coherent state. Moreover, the circular state is
symmetric, ${\6l\,l}|\textbf{r}|{l\,l\9}=0$, while the extreme
Stark state state is not. This can be seen from
\cite{wybourne}:
\beq
{\6nlm}|\textbf{r}|{nlm\9}=\mbox{$2\over
3$}{\6nlm}|\textbf{K}|{nlm\9}.
\eeq
Let us examine which one of these states gives better results when
used by Alice to transmit the directions of her $z$-axis. The
overlap between two angular momentum coherent states
is~\cite{peresbook}
\beq
|{\6\textbf{u}_1}|{\textbf{u}_2\9}|^2=
\cos^{4j}(\chi/2),
\label{overlapj}
\eeq
where $\chi$ is the angle between the directions of $\bu_1$ and
$\bu_2$. It is noteworthy that the overlap between two extreme
Stark states is the same, as we will see shortly. First, a
rotation of the $|{K, \textbf{z}\9}$ state by angles
$(\theta\phi)$ gives
\beq
|{K,\bu_{\theta\phi}\9}= e^{-iL_z\phi} e^{-iL_y\theta}|{K,\bz\9},
\eeq
where again the operator $e^{-iL_z\phi}e^{-iL_y\theta}$ performs
an active rotation of the vector $|{K,
\textbf{z}\9}$. Using (\ref{LA}) we have
\begin{eqnarray}
\!\!\!\!|{K,\bu_{\theta\phi}\9}\!\!\!&=&\!\!
e^{-i(J_{1z}+J_{2z})\phi}e^{-i(J_{1y}+J_{2y})\theta}|{-\textbf{z}\9}
\otimes |{\textbf{z}\9},\\
\!\!\!\!\!\!&=&\!\!e^{-iJ_{1z}\phi}e^{-iJ_{1y}\theta}|{
-\textbf{z}\9}\otimes
e^{-iJ_{2z}\phi}e^{-iJ_{2y}\theta} |{\textbf{z}\9},
\label{starkgeneral}
\end{eqnarray}
owing to Eqs. (\ref{com1}) and (\ref{com2}). Thus the rotated
extreme Stark state is just
\beq
|K,\bu_{\theta\phi}\9 =|-\bu_{\theta\phi}\9\otimes
|\bu_{\theta\phi}\9,
\eeq
where the SO(3) coherent states $|\bu_{\theta\phi}\9$ are defined
as in Eq.~(\ref{ji}). This Stark state is an eigenstate of
$\textbf{u}\cdot\textbf{K}$ with the maximal eigenvalue $n-1$, and
it satisfies $\bk=\bu$, as can be seen from Eq.~(\ref{kl}). The
overlap between two such states,
$|{\6K,\textbf{u}'}|{K,\textbf{u}''\9}|^2$, is
\beq
|{\6 -\textbf{u}'}|{ -\textbf{u}''\9}|^2\;|{\6
\textbf{u}'}|{\textbf{u}''\9}|^2,
\eeq
which by using (\ref{overlapj}) is just:
\beq
\cos^{4j_1}(\chi/2)\cos^{4j_2}(\chi/2)=\cos^{4(n-1)}(\chi/2),
\label{overlapa}
\eeq
where $\chi$ is the angle between the vectors $\bu'$ and $\bu''$.

Such a simple expression cannot hold for the overlap of two
generic elliptic states whose eccentricities are not 0 or~1. Let a
generic elliptic state
\beq
|{\textbf{u}_1 \textbf{u}_2\9} = |{\textbf{u}_1\9} \otimes
|{\textbf{u}_2\9},
\eeq
be an elliptic state with eccentricity $0<e<1$. Unlike the $e=1$
and $e=0$ cases, this state does not define one direction, but two
independent ones $\bu_1$ and $\bu_2$. If it is rotated by Euler
angles $\alpha\beta\gamma$ the result is

\beq
e^{-iL_{z}\alpha}e^{-iL_{y}\beta}e^{-iL_{z}\gamma}|{\bf{u}}_1\,\textbf{u}_2\9
= U_1|\textbf{u}_1\9
\otimes U_2|\textbf{u}_2\9.
\eeq
where
\beq
U_1=e^{-iJ_{1z}\alpha}e^{-iJ_{1y}\beta}e^{-iJ_{1z}\gamma},
\label{Uj}
\eeq
and likewise for $U_2$. To obtain this result we have
used~(\ref{LA}) and the commutation relations (\ref{com1}) and
(\ref{com2}). The rotation
$e^{-iL_{z}\alpha}e^{-iL_{y}\beta}e^{-iL_{z}\gamma}$ opens an
angle $\chi_1$ between the classical vectors $\textbf{u}_1$ and
$R(\alpha\beta\gamma)\textbf{u}_1$, and an angle $\chi_2$ (which
is generally different from $\chi_1$) between $\bu_2$ and
$R(\alpha\beta\gamma)\bu_2$. Here $R(\alpha\beta\gamma)$ denotes
the classical rotation matrix
\cite{gold}. It follows that

\beq \6\bu_1\bu_2|e^{-iL_z\alpha}e^{-iL_y\beta}
e^{-iL_z\gamma}|\bu_1\bu_2\9=\left(
\cos{\chi_1\over2}\cos{\chi_2\over2}\right)^{2(n-1)}.\eeq
Generally, both $\chi_1$ and $\chi_2$ are different from the angle
between the directions $\textbf{k}$ and
$\textbf{k}'=R(\alpha\beta\gamma)\bk$, or between the directions
$\bl$ and $\bl'=R(\alpha\beta\gamma)\bl$.

We now calculate the transmission fidelity for the case where
Alice sends an extreme Stark state $|{K,\textbf{z}\9}$. Since
$|A\9$ contains only $m=0$ terms, so does Bob's fiducial vector
\beq
b_{lm} =a_{l0}(|a_{l0}|^2)^{-1/2}\delta_{m 0}.
\eeq
We thus have
\beq
b_{lm} = \delta_{m 0}\,(a_{l 0}/|a_{l 0}|),
\label{bobstark}
\eeq
\beq |{ B\9} =
\sum_{l=0}^{n-1}\sqrt{2l+1}\;(a_{l 0}/|a_{l 0}|)\,|{l\,0\9}.
\eeq
In order to determine $\6\cos\omega_z\9$ in
Eq.~(\ref{expectation}), we note that
\bea
{\6 A}|U(\alpha\beta \gamma)|{B\9}
&=&\sum_{l=0}^{n-1}\sqrt{2l+1}\,a_{l0}^{*}b_{l0}{\6 l\,0}|
\cD^{\,(l)}(\alpha\beta \gamma)|{l 0\9},\nonumber\\
    &=&\sum_{l=0}^{n-1}\sqrt{2l+1}\,|a_{l0}|d_{00}^{\,(l)}(\beta).
\eea
We insert this expression into~(\ref{expectation}). The result,
obtained by using Eqs.~(19)--(21) of ref.
\cite{peres1}, is
\beq
\6\cos\omega_z\9=\sum_{k l}\,A_{l k}\,|a_{l 0}\,a_{k 0}|,
\label{fidelm0}
\eeq
where $A_{l k}$ is a real symmetric matrix whose non-vanishing
elements are
\beq
A_{l,l-1}=A_{l-1,l}=l/\sqrt{4l^2-1},
\label{calc4}
\eeq
and $a_{l\,0}=C^{jj\,l}_{-j j 0}$. The results are summarized in
Fig.~1, in which the mean square error is plotted versus $n$. The
$|{K,\textbf{z}\9}$ state gives better fidelity than the circular
state $|{ll\9}$, but for $n>3$ its fidelity is substantially less
than optimal \cite{peres1,catalans1} and goes asymptotically to
$1/(4n-2)$. This raises the question whether it is possible to
build a ``natural" POVM by setting Bob's vector to $|{
B\9}=\sqrt{N}|{K,\textbf{z}\9}$, so that POVM elements are
\beq
N|{K, \textbf{u}_{\theta\phi}\9}{\6K,\textbf{u}_{\theta\phi}}|,
\eeq
where $|{K, \bu_{\theta\phi}\9}$ was defined in
Eq.~(\ref{starkgeneral}) and $N$ is a normalization factor.
Unfortunately, $|{K,
\textbf{z}\9}$ contains a superposition of all values of
$l$, as can be seen from (\ref{extreme}). Thus $|{K,
\textbf{z}\9}$ does not belong to one irreducible subspace of the
representation of the SO(3) rotation group. As a result, the
operator
\beq B=\int
d_{\theta\phi}|{K,
\textbf{u}_{\theta\phi}\9}{\6K, \textbf{u}_{\theta\phi}}|
\eeq
is not proportional to the identity, but is a block-diagonal
matrix with different blocks for each irreducible representation
of the rotation group. Moreover, the resulting POVM includes an
element which corresponds to the absence of any answer, thus
reducing fidelity. A natural POVM which uses the SO(4) group will
be discussed in Sec.~IV.

The direction of the minor axis of a classical nondegenerate
ellipse is that of $\textbf{L}\times\textbf{K}$. A quantum ellipse
also has this property. Taking Alice's state as a quantum ellipse
with eccentricity $0<e<1$, with both $\bk$ and $\bl$ lying in the
$xy$ plane so that $\bw=\bz$, the resulting fidelity can be
compared  with the cases where $\bk$ or $\bl$ points along the
$z$-axis and the eccentricity of the ellipse is 0 or 1,
respectively. The fidelity for transmission using the semi-minor
axis reaches a maximum at eccentricity of about $e=0.7$ (a
different eccentricity for each value of $n$) . A comparison of
the mean square error for using the three options is given in
Table 1.

\subsection{Comparison between elliptic wave functions and optimal wave
functions}

We shall now compare the extreme Stark state with Alice's optimal
vector for the transmission of one axis as calculated in
\cite{peres1}. They are both eigenstates of $L_z$ with
$m=0$, and since Eq.~(\ref{fidelm0}) holds, we will present them
in the notation: $(|a_{00}|,|a_{10}|,|a_{20}|,...,|a_{n-1,0}|)$,
where $a_{l\,0}=C^{jj\,l}_{-jj0}$ as before. For $n=3$ we have
\beq
|{K,
\bz\9}=({\mbox{$1\over\sqrt{3}$}},{\mbox{$1\over\sqrt{2}$}},{\mbox{$1\over\sqrt{6}$}}),
\eeq
while Alice's optimal state is
\beq
|{ A_{{\rm opt}}\9}
=({\mbox{$\sqrt{5}\over3\sqrt{2}$}},{\mbox{$1\over\sqrt{2}$}},{\mbox{$\sqrt{2}\over3$}}).
\eeq
Thus for $n=3$ the overlap between the extreme Stark state and the
optimal state is
\beq
|{\6K, \bz}|{ A_{{\rm opt}}\9}|^2 = 0.993491.
\eeq
Both states give almost the same fidelity for transmission of one
axis. For higher values of $n$, they become more and more
different. For $n=10$ the overlap is
\beq
|{\6K, \bz}|{ A_{{\rm opt}}\9}|^2 = 0.76406.
\eeq
The various components are given in Table 2. We see that the
extreme Stark state has coefficients peaked at lower values of $l$
than the optimal state.

\section{ Transmission of Two Axes}
Alice now wants to transmit a Cartesian frame by indicating the
directions of two axes, the third one being inferred from them.
Which elliptic state is optimal? Obviously, states with $e=0$ and
$e=1$ will not do in this case, since they define only one
direction. We have to find the optimal eccentricity. Let
\beq
\Delta K_\perp =\sqrt{(\Delta K_\ell)^2+(\Delta K_w)^2},
\eeq
\beq
\Delta L_\perp =\sqrt{(\Delta L_k)^2+(\Delta L_w)^2},
\eeq
where
\beq
\Delta K_\ell = \sqrt {\6K_\ell^2\9 - \6K_\ell\9^2}=\sqrt{\6K_\ell^2\9},
\eeq
owing to Eq.~(\ref{experp}). We define similar expressions for the
other components. When we want to transmit $\bk$, namely the
direction of the classical LRL vector, then a smaller ${\Delta
K_\perp}/{\6K_k\9}$ improves the fidelity. A similar argument
holds for the transmission of $\bl$. Thus when transmitting two
axes, a heuristic guideline is to look for states that satisfy
\beq
\frac{\Delta K_\perp}{\6K_k\9}\approx\frac{\Delta L_\perp}{\6L_l\9}.
\eeq
A straightforward calculation \cite{delande} gives
\beq
{\Delta K_w} = {\Delta L_w} =\sqrt{\half(n-1)},
\eeq
\beq
{\Delta K_\ell} = \sqrt{\half(n-1)}\sin\zeta,
\eeq
\beq
{\Delta L_k} =
\sqrt{\half(n-1)}\cos\zeta.
\eeq
Together with Eqs.~(\ref{expK}) and (\ref{expL}), this gives an
equation for the eccentricity,
\beq
\frac{\sqrt{1+ \sin^2\zeta}}{\sqrt{2(n-1)}\sin\zeta}\approx
\frac{\sqrt{1+\mathstrut\cos^2\zeta}}{\sqrt{2(n-1)}\cos\zeta}.
\eeq
Therefore we expect that the optimal eccentricity is approximately
\beq
e=\sin{\zeta}=\cos\zeta=1/\sqrt{2}.
\label{opte}
\eeq
More accurate numerical results are given below.

We now evaluate the fidelity for the transmission of two axes.
Alice uses an elliptic state with $\bk = {\bf x}$ and $\bl = {\bf
y}$ (the unit vectors in the $x$ and $y$ directions respectively).
The eccentricity $e=\sin\zeta$ has to be optimized. Recall the
$\zeta$ is defined to be half the angle between $\bu_1$ and
$\bu_2$. The definitions of $\bk$ and $\bl$ are given in
Eq.~(\ref{kl}). Thus in order to meet the above requirements we
set in $|{A\9}=|{\bu_{\theta_1\phi_1},
\bu_{\theta_2\phi_2}}\9$ the parameters $\theta_1=\theta_2=\pi/2$, and
\beq
\phi_1=\half\pi-\zeta, \qquad\phi_2={\mbox{$3\over2$}}\pi-\zeta.
\label{alice2}
\eeq
Fidelities now must be defined for each one of the axes. Note that
$\cos\omega_k$ (for the {\it k\/}th axis) is given by the
corresponding diagonal element of the orthogonal (classical)
rotation matrix. For the transmission of the $x$ and $y$ axes, we
thus need \cite{gold}
\beq \6\cos\omega_x + \cos\omega_y\9 = \6(1 + \cos\beta)(\cos(\alpha
+ \gamma)\9. \eeq

We expand $|{A\9}$ and $|{B\9}$ as in (\ref{alicegeneral}) and
(\ref{bobgeneral}). Bob's optimal fiducial vector is still given
by (\ref{bob}), and $\6\cos\omega_x +
\cos\omega_y\9$ is calculated using equations (23)--(26) of
ref.~\cite{peres2}. The mean square error per axis is plotted in
Fig.~3 for $n=5,10,$ and $20$. The error is minimal at $e\approx
0.708$ for $n=5$, at $e\approx0.704$ for $n=10$, and
$e\approx0.674$ for $n=20$.

Note that the shape of the curve flattens with increasing $n$, so
that the minimum is hard to find numerically. The intuitive
explanation is that in the limit of large $n$, as if Alice were to
transmit a ``classical atom," {\it i.e.}, a classical two body
Kepler system, then the direction of the classical angular
momentum and LRL vectors could be found irrespective of the
eccentricity. Therefore, the transmission accuracy would be the
same for any eccentricity that is not close to zero or one.

The deviation of the optimum from $e=1/\sqrt{2}$ was expected,
since transmission of the $\bk$ direction ($e=1$) achieved higher
fidelity than the transmission of the $\bl$ direction ($e=0$).
Thus the ellipse with optimal eccentricity for transmission of two
axes is biased to give ${\Delta L_\perp}/{L}<{\Delta K_\perp}/{K}$
in order to compensate the difference and make the contribution to
the error from the $\bk$ direction about equal to that from the
$\bl$ direction.

Elliptic states give results very close to the optimal ones. The
mean square error for transmission of two axes by elliptic states
with optimal eccentricity is compared to the optimal results
\cite{peres2} in Table III.

\section{POVM for SO(4)}
We now construct a POVM based on the SO(4) group and use it in
order to transmit two axes. This POVM is naturally built with the
SO(4) coherent states which are, as we have seen, direct products
of two SO(3) coherent states. We shall use for each one of the
SO(3) subspaces the notation
\beq
|{\psi\theta\phi\9}_{\bu}=\sqrt{2j+1}\;U(\psi\theta\phi)|{\bu\9},
\eeq
and
\beq
dE(\psi\theta\phi)=d_{\psi\theta\phi} |{\psi\theta\phi\9}_{\bu}
{\6\psi\theta\phi}|_{\bu},
\eeq
where $\bu$ labels the direction to be transmitted, and
$d_{\psi\theta\phi}=\sin\theta d\theta d\phi d\psi /{8\pi^2}$ as
in Eq.~(\ref{povm}). By applying Schur's lemma to each of the
SO(3) subspaces we have
\beq \int\!\!\!\int dE_1(\psi_1\theta_1\phi_1)\otimes
dE_2(\psi_2\theta_2\phi_2)=\1_1\otimes\1_2=\1.
\eeq

We are now ready to discuss the transmission of Alice's $x$ and
$y$ axes by means of an elliptic state. We take
\beq
|{A \9} = |{\bx \by \9} = |{\bx\9}\otimes  |{\by\9}.
\label{aliceso4}
\eeq
This equation was written in Alice's notation. We also define a
fiducial vector for Bob
\beq |{B \9} = (2j+1)|{\bx\9}\otimes
{|\by\9},
\label{bobso4}
\eeq
written in Bob's notations. Thus the POVM element is constructed
from the vector
\bea
|{\psi_1\theta_1\phi_1\9}_\bx
\otimes|{\psi_2\theta_2\phi_2\9}_\by
=(2j+1)(U_1\otimes U_2)|{B
\9}.
\eea

The result of Bob's measurement consists of two sets of Euler
angles, $\psi_1\theta_1\phi_1$ and $\psi_2\theta_2\phi_2$. The
first one gives Bob's estimate of the active rotation needed to
bring his $x$-axis to Alice's $x$-axis. Likewise, the second set
gives Bob's estimate of the active rotation needed to bring his
$y$-axis to Alice's $y$-axis. The detection probability of these
sets of angles is
\beq
dP(\psi_1...\phi_2)=
d_{\psi_1\theta_1\phi_1}d_{\psi_2\theta_2\phi_2} |{\6A}|U_1
\otimes U_2|{B\9}|^2.
\label{prob2notation}
\eeq
Recall that Eq.~(\ref{aliceso4}) was written in Alice's notations,
while (\ref{bobso4}) is in Bob's notations. To compute the result
explicitly, we need a uniform system of notations. For this we
introduce, as in \cite{peres2}, the Euler angles $\xi \eta
\zeta$ that rotate Bob's $xyz$ axes into Alice's axes. (The Euler
angle $\zeta$ should not be confused with the eccentricity
parameter introduced in Sec. II.) The unitary operator $U(\xi
\eta \zeta)$ represents an active transformation of Bob's state vectors
to the corresponding state vectors of Alice's system. Therefore,
$U(\xi
\eta
\zeta)$ is also the passive transformation from Alice's notations
to Bob's notations. Written in Bob's notations, Alice's vector
$|{A \9}$ becomes $U(\xi \eta
\zeta)|{A \9}$ so that in Eq.~(\ref{prob2notation}), ${\6 A}|$
becomes ${\6 A}|U(\xi \eta
\zeta)^\dag$. Owing to the commutation relations
(\ref{com1}) and (\ref{com2})
\bea
U(\xi \eta \zeta)&=&
e^{-iL_{z}\xi}e^{-iL_{y}\eta}e^{-iL_{z}\zeta},\\
&=&U_1(\xi \eta
\zeta)\otimes U_2(\xi \eta \zeta),
\eea
where again $U_1$ and $U_2$ are defined as in Eq.~(\ref{Uj}). We
thus have
\beq U(\xi \eta \zeta)|{\bu_1\bu_2\9}=U_1(\xi \eta
\zeta)|{\bu_1\9}\otimes U_2(\xi \eta \zeta)|{\bu_2\9}.
\eeq
 Let us therefore define
\beq
U_1(\alpha_1 \beta_1 \gamma_1) = U_1^\dag(\xi \eta
\zeta)\,
U_1(\psi_1\theta_1\phi_1),
\label{U1}
\eeq
and
\beq
U_2(\alpha_2 \beta_2 \gamma_2) = U_2^\dag(\xi \eta
\zeta)\,U_2(\psi_2\theta_2\phi_2).
\label{U2}
\eeq
We shall henceforth use the left hand sides of Eqs.~(\ref{U1}) and
(\ref{U2}) as the new definitions of the symbols $U_1$ and $U_2$.
As before, the Euler angles $\alpha_1
\beta_1
\gamma_1$ have the effect of rotating Bob's $x$-axis into his
estimate of Alice's $x$-axis and then rotating back the result by
the true rotation from Alice's to Bob's frame. The action of the
Euler angles $\alpha_2\beta_2\gamma_2$ is similar for the
$y$-axis. Thus the Euler angles $\alpha_i\beta_i \gamma_i$
indicate Bob's measurement error, and the probability of that
error is
\beq
dP(\alpha_1...\gamma_2)= d_{\alpha_1 \beta_1
\gamma_1}d_{\alpha_2 \beta_2 \gamma_2}|{\6 A}|U_1\otimes
U_2|{B\9}|^2.  \label{prob} \eeq Note the similarity with
Eq.~(\ref{prob2notation}). The difference is that
(\ref{prob2notation}) referred to the probability of {\it
detection} of a particular set of Euler angles, while (\ref{prob})
gives the probability of {\it error} in that detection.

The transmission mean square error {\it per axis} is, as in
Eq.~(\ref{msecirc}),

\beq R = \qtr(1-\cos \omega_{x})+\qtr(1-\cos \omega_{y}), \eeq
where $\omega_x$ and $\omega_y$ are the angles between the true
and estimated directions of the $x$-axis and $y$-axis,
respectively. Since Bob infers the direction of the $x$-axis from
the angles $\psi_1\theta_1\phi_1$, the value of $\cos\omega_{x}$
depends only on $\alpha_1\beta_1\gamma_1$. Likewise, the value of
$\cos\omega_{y}$ depends only on the angles
$\alpha_2\beta_2\gamma_2$. We have
\beq
\6\cos\omega_{x}\9=\int d_{\alpha_1 \beta_1
\gamma_1} |{\6 \bx}| U_1|{\bx\9}|^2\cos
\omega_{x},
\label{wxso4}
\eeq
where we have used (\ref{prob}) and Schur's lemma for the second
set of angles, namely
\beq
(2j+1)\int d_{\alpha_2 \beta_2 \gamma_2}
U_2|{\by\9}{\6\by}|U_2^\dag = \1_2.
\eeq
The evaluation of Eq.~(\ref{wxso4}) is identical to the one
performed in Eq.~(\ref{wzcircular}),  with $n$ replaced by
$\half(n+1)$ everywhere, and we get
\beq
\6\cos\omega_{x}\9 = (n-1)/(n+1).
\eeq
Likewise
\beq
\6\cos\omega_{y}\9 = (n-1)/(n+1).
\eeq
Thus the infidelity (mean square error) per axis is
\beq
\qtr(1-\6\cos
\omega_{x}\9)+\qtr(1-\6\cos\omega_{y}\9) = 1/(n+1)
\eeq
In ref.~\cite{peres2} it was found that the {\it optimal} POVM
(not restricted to elliptic states) for transmission of two axes
using a hydrogen atom is of the form given by Eq.~(\ref{povm}). It
was shown that using this POVM, the infidelity per axis for
Alice's optimal signal approaches $1/(3n)$ asymptotically. Using
SO(4) instead of SO(3) as in~\cite{peres2}, we obtain an
infidelity per axis that is exactly $1/(n+1)$ for all values
of~$n$. As shown in the Appendix, an adjustment procedure to
obtain to orthogonal axes will further decrease the mean square
error by a factor which for large values of $n$ tends to $3/4$.

The SO(4) POVM also enables the transmission of two directions
which are not ortho\-gonal, by means of a single hydrogen atom in
an elliptic state. To transmit the directions of two general unit
vectors $\bv_1$ and $\bv_2$, Alice's prepares the elliptic state
\beq
|{A \9} = |{\bv_1\bv_2 \9} = |{\bv_1\9}\otimes  |{\bv_2\9},
\eeq
(in her notations) while Bob's vector is (in his notations)
\beq |{B \9} = (2j+1)|{\bv_1\9}\otimes
{|\bv_2\9}.
\eeq
As before, the infidelity for each direction is $1/(n+1)$. It
should be noted that transmission of two non-orthogonal directions
with one hydrogen atom is not possible with the SO(3) POVM.

\section{Summary and concluding remarks}
We have shown how elliptic Rydberg states can transfer information
on the orientation of one direction, or more generally that of a
Cartesian frame. For increasing values of $n$, the fidelity
obtained for a single direction falls rapidly below the optimal
ones. However, for a Cartesian frame the results are very close to
the optimal ones. Furthermore, elliptic states have the advantage
of being experimentally accessible, while preparation of the
optimal states seems much more difficult. Note that we have
assumed Alice and Bob have the same chirality. If their
chiralities are opposite, then when angular momenta are used for
the transmission, the direction inferred by Bob should be reversed
(because directions are polar vectors while angular momentum is an
axial vector). However, the LRL vector is also a polar vector,
thus even if Bob and Alice have opposite chiralities, the
direction inferred by Bob is correct. We have also shown how
elliptic Rydberg states can be prepared to encode two arbitrary
directions, when the measurement is based on the SO(4) rotation
group.

\bigskip Work by AP was supported by the Gerard Swope Fund and the
Fund for Promotion of Research. Work by NHL was supported by a
grant from the Technion Graduate School.

%\begin{center}
%\appendix{Reducing errors with an orthogonalization procedure}
%\end{center}
\appendix

\section{Reduction of errors by orthogonalization}
As we have seen in Sec.~V, Bob's estimates of Alice's $x$ and $y$
axes may not be exactly orthogonal. The probability for the
estimate of the $x$ axis to have an angular error $\omega_x$, as
can be seen from Eq.~(\ref{overlapj}), is
\beq
\rho(\omega_x)\propto\cos^{2n-2}(\omega_x/2),
\label{probomega}
\eeq
and likewise for the $y$ axis. Thus for large values of $n$, the
error probability distribution will be highly peaked. We now
calculate the gain in fidelity achieved if Bob performs a simple
orthogonalization of his two estimates $\hbr_x$ and $\hbr_y$, by
rotating the two vectors in their plane by the same angle, so that
they become orthogonal.

Let us define two pairs of spherical angles that give the position
of the estimated directions with respect to the (unknown) true
axes. These positions are given by
\beq
\hbr_x=(\sin\theta_1\cos\phi_1, \sin\theta_1\sin\phi_1,
\cos\theta_1),
\label{r1}
\eeq
and
\beq
\hbr_y=(\sin\theta_2\cos\phi_2, \sin\theta_2\sin\phi_2,
\cos\theta_2).
\label{r2}
\eeq
The probability distributions will be denoted
$\rho_i(\theta_i,\phi_i)$. In the limit of large $n$, the
deviation angles $\omega_x$ and $\omega_y$ are small. Hence the
distribution are centered as
\beq
\rho_x=\rho(\theta_1-\half\pi,\phi_1),\qquad
\eeq
\beq
\rho_y=\rho(\theta_2-\half\pi,\phi_2-\half\pi),
\eeq
where $\rho(\xi,\mu)$ is peaked around $(0,0)$. Here we used the
fact that the SO(4) POVM gives probabilities of error for each
axis which are identical and independent. Define new variables
\beq
\tilde{\theta}_i=\theta_i-\half\pi,
\label{tal}
\eeq
\beq
\tilde{\phi}_2=\phi_2-\half\pi.
\label{phitilde}
\eeq
The deviation angles are given by $\cos\omega_x=\hbr_x\cdot\hbx$
and $\cos\omega_y=\hbr_y\cdot\hby$, namely
\beq
\cos\omega_x=\sin\theta_1\cos\phi_1\approx
1-\half\tal_1^2-\half\phi_1^2,
\label{coswx}
\eeq
and
\beq
\cos\omega_y=\sin\theta_2\cos\phi_2\approx
1-\half\tilde{\theta}_2^2-\half\tilde{\phi}_2^2.
\eeq

Let $g$ denote the infidelity per axis before the adjustment. The
infidelities for both axes are equal, thus
\bea
\lefteqn{g\equiv \half (1- \6\cos\omega_x\9),}\\
&& \phantom{g}\approx\qtr\int
\left(\tal_1^2+\phi_1^2 \right) d\rho_i\equiv
\qtr\6\tal_1^2+\phi_1^2\9,
\label{fx}
\eea
where
\beq
d\rho_i=\rho(\tal_i,\phi_i)\sin\tal_i\,d\tal_i\,d\phi_i,
\eeq
fullfills
\beq
\int d\rho_i=1.
\eeq
Equivalently, we can write the infidelity in terms of $\tal_2$ and
$\tilde{\phi}_2$ as
\beq
g\approx\qtr\int \left(\tilde{\theta}_2^2+\tilde{\phi}_2^2\right)
d\rho_i\equiv \qtr\6\tilde{\theta}_2^2+\tilde{\phi}_2^2\9.
\label{fy}
\eeq
In {\em first} order we have, by combining (\ref{r1}) and
(\ref{r2}) with the definitions (\ref{tal}) and (\ref{phitilde}),
\beq
\hbr_x\approx(1,\phi_1,-\tal_1),\qquad
\hbr_y\approx(-\teb_2,1,-\tal_2),
\label{rapprox}
\eeq
and the angle $\Omega$ between them is given by
\beq
\cos\Omega=\hbr_x\cdot\hbr_y\approx\phi_1-\teb_2 \label{omega}.
\eeq
The bisector of $\hbr_x$ and $\hbr_y$ is given by the {\em unit}
vector $\hbb={(\hbr_1+\hbr_2)}/|\hbr_1+\hbr_2|$.
Using~(\ref{rapprox}) and keeping only first order terms we have
\beq
\hbb\approx\left[
1-\half(\phi_1+\teb_2),1+\half(\phi_1+\teb_2),-\tal_1-\tal_2
\right]/\sqrt{2},
\eeq
where we used
\beq
|\hbr_1+\hbr_2|\approx\sqrt{2}(1+\half\phi_1-\half\teb_2).
\eeq
We can also express the bisector $\hbb$ in terms of its spherical
angles which we shall denote by $(\tau,\varphi)$. Since the errors
are small, we have $\varphi\approx\qtr\pi$, and it is convenient
to define
\beq
\tap=\varphi-\qtr\pi.
\eeq
Comparison of the two expressions for $\hbb$ gives
\beq
\xi=\half\pi+\sqrt{\half}(\tal_1+\tal_2),\qquad
\tap=\half(\phi_1+\teb_2).
\eeq

In first order, as Eq.~(\ref{omega}) shows, the orthogonalization
consists in changing the angles $\phi_i$ irrespective of
$\theta_i$, without changing the $\theta_i$ themselves. Hence, in
first order, the procedure defines
\beq
\phi'_1=\varphi-\qtr\pi, \qquad \phi'_2=\varphi+\qtr\pi,
\eeq
{\it i.e.,}
\beq
\phi'_1=\teb'_2=\tap=\half(\phi_1+\teb_2),
\eeq
where again $\teb'_2 = \phi'_2-\half\pi$. The change in $\tal_i$
is of higher order, $\tal'_i=\tal_i+O(\theta^2,\phi^2)$. The new
infidelity per axis $g^{\rm new}$ is
\beq
g^{\rm
new}=\qtr\6\phi_1^{'2}+\tilde{\theta}_1^2\9=\qtr\6\teb_2^{'2}+\tilde{\theta}_2^2\9.
\eeq

Returning to (\ref{fx}) and (\ref{fy}), consider the integrals
over $\phi_i$. Define
\beq
\rho_{\phi}(\phi)\equiv \int \rho(\theta,\phi) \sin\theta d\theta
\eeq
Keeping in mind that the distributions for $\phi_1$ and $\teb_2$
are identical, the $\phi$-part of the infidelity per axis {\it
before} the adjustment is
\bea
\lefteqn{g_\phi=\qtr\6\phi_1^2\9=\qtr\6\teb_2^2\9,} \label{equal}\\
&&\phantom{g_\phi}=\int\phi_1^2\,\rho_{\phi}(\phi_1)
\rho_{\phi}(\teb_2)d\phi_1d\teb_2/4,\\
&& \phantom{g_\phi}=\int\teb_2^2\,\rho_{\phi}(\teb_2)
\rho_{\phi}(\teb_2)d\phi_1d\teb_2/4.
\eea

The $\phi$-parts of the infidelities {\it after} the adjustment,
denoted by $g^{\rm new}_\phi$, are
\beq
g^{\rm new}_\phi = \qtr\6\phi_1^{'2}\9=\qtr\6\teb_2^{'2}\9=\ses\6
\phi_1^{2} +2\phi_1\teb_2 + \teb_2^{2}\9.
\eeq
The functions $\rho_\phi(\phi_1)$ and $\rho_\phi(\teb_2)$ are
even, because the probability distribution $\rho$ depends only on
the angles $\omega_x$ or $\omega_y$, which are independent of the
sign of $\phi_1$ and $\teb_2$ . Thus
\beq
\6\phi_1\teb_2\9=\int \phi_1\teb_2
\,\rho_{\phi}(\phi_1)\,\rho_{\phi}(\teb_2)\,d\phi_1
\,d\teb_2=0.
\eeq
With the help of Eq.~(\ref{equal}) we obtain
\bea
g^{\rm new}_\phi = \ses \6 \phi_1^{2} +
\teb_2^{2}\9 = \mbox{$1\over8$} \6 \phi_1^2\9.
\eea

Thus the $\phi$-parts of the infidelity are halved,
\beq
g^{\rm new}_\phi=\half g_\phi.
\eeq
As already stated, the angles $\tilde{\theta_i}$ are unchanged in
first order. Since the probability function
$\rho_x(\phi_1,\tal_1)$ depends only on the angle $\omega_x=\hbr_x
\cdot \hbx$, it is symmetric with respect to rotations around the
$x$ axis. A similar argument holds for the $y$-axis. Thus
\beq
\6\tal_1^2\9=\6\phi_1^2\9=\6\tal_2^2\9=\6\phi_2^2\9,
\eeq
and we have finally
\beq
g^{\rm new}=\mbox{$3\over4$}g.
\eeq

\newpage

\begin{table}
\begin{center}
\begin{tabular}{clll}
$n$ & $\bz=\bw$ & $\bz=\bl$  &  $\bz=\bk$ \\
\hline
 5 & $e=0.6963$ & $e=0$ & $e = 1$ \\
   & $\eta=0.193967$  & $\eta=0.1$ & $\eta=0.0573645$\\ \hline
 10 & $e=0.701261$ & $e=0$ & $e=1$ \\
   & $\eta=0.0861934$ &  $\eta=0.05$ & $\eta=0.0264067$\\
\end{tabular}
\caption{Eccentricities $e$ and mean square errors $\eta$ for
transmission of a single direction using $\bz=\bw$, $\bz=\bl$, or
$\bz=\bk$, for $n=5$ or $10$.}
\end{center}
\end{table}

\bigskip

\begin{table}
\begin{center}
\begin{tabular}{lcccccccccc}
$l$ & 0 & 1 & 2& 3 & 4 & 5 & 6 & 7 & 8 & 9 \\ \hline $|{K,\bz\9}$&
0.3162 & 0.4954 & 0.5222 & 0.4534 & 0.3365 & 0.2148 & 0.1167 &
0.0526 & 0.0186 & 0.0045 \\ \hline  Optimal& 0.1825 & 0.3079 &
0.3767 &
0.4098 & 0.4130 & 0.3894 & 0.3422 & 0.2751 & 0.1923 & 0.0989 \\
\end{tabular}
\caption{Coefficients $|a_{l0}|$ for Alice's optimal state and for
the extreme Stark state when $n=10$.}
\end{center}
\end{table}

\bigskip
\begin{table}
\begin{center}
\begin{tabular}{llccc}
$n$ &  &elliptic & & optimal  \\
\hline
 5 & & $\eta = 0.14765$ &  &$\eta = 0.14465$\\  \hline
 10 & & $\eta=0.06822$ &   &$\eta=0.06793$ \\ \hline
 20 & &$\eta=0.03190$ &  &$\eta=0.03088$
\end{tabular}
\caption{Mean square error $\eta$ for
transmission of two axes by an elliptic state with optimal
eccentricity, and by the optimal method \cite{peres2} for $n=5,
10,$ and $20$.}
\end{center}
\end{table}

\begin{figure}
\begin{center}
\epsfxsize=0.48\textwidth
\centerline{\epsffile{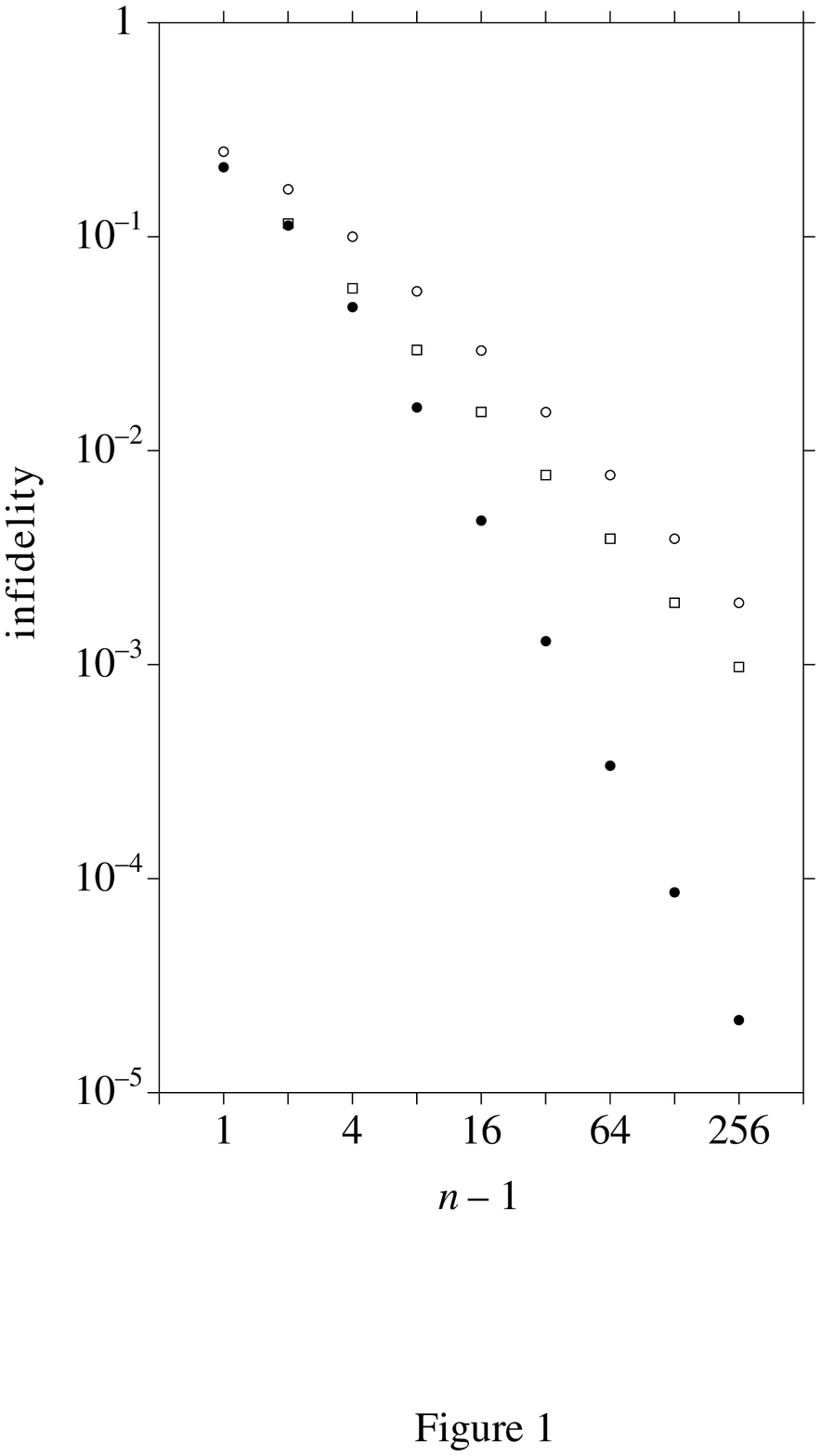}} \vspace*{-0.1cm}
\caption{Mean square error as a
function of $n$ for the transmission of a single axis using the
circular state (open circles), the extreme Stark state (squares),
and the optimal state (closed circles).}
\end{center}
\end{figure}

\begin{figure}
\begin{center}
\epsfxsize=0.48\textwidth
\centerline{\epsffile{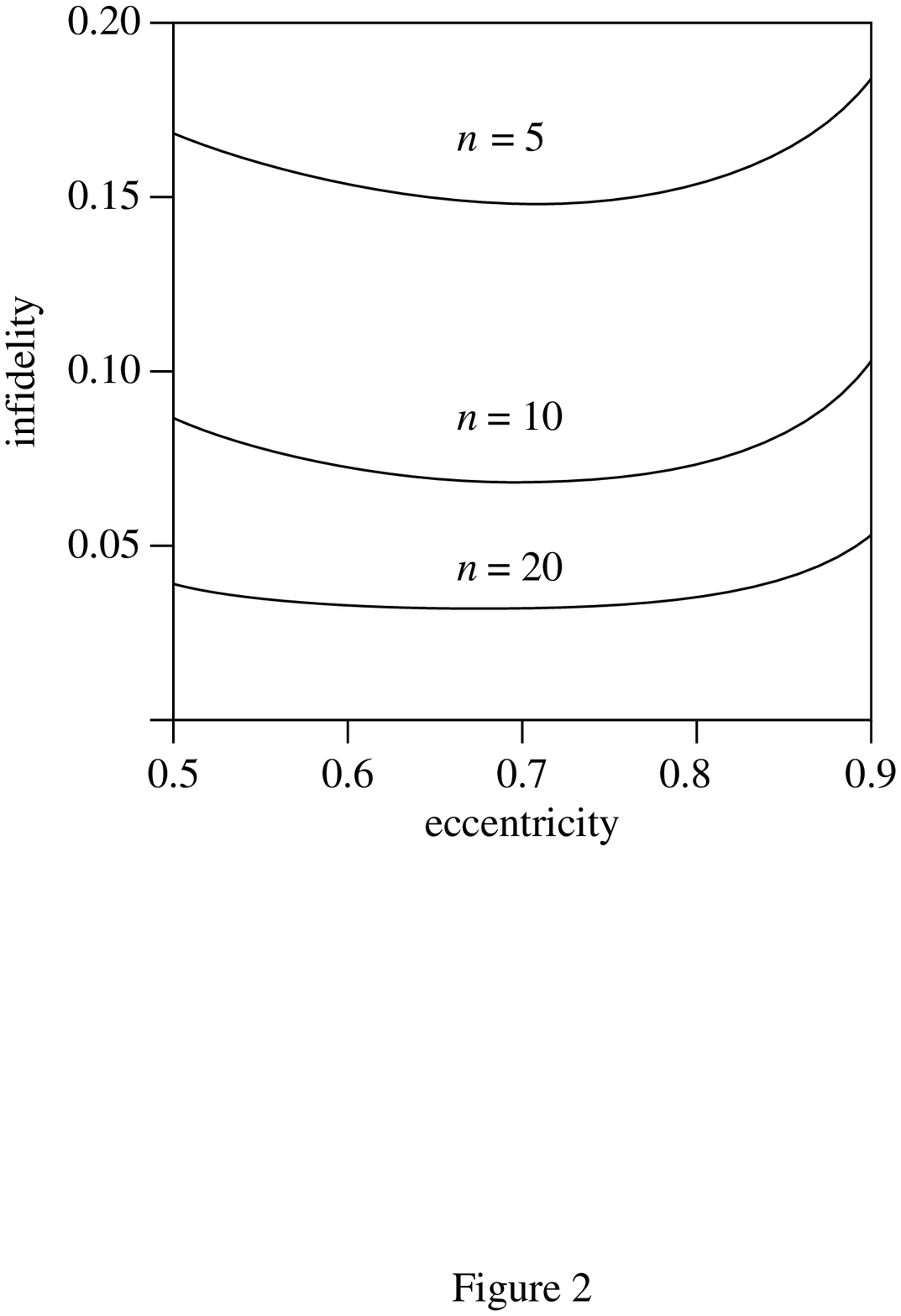}} \vspace*{-0.1cm}
\caption{Mean square error (per axis) as a
function of eccentricity for $n$=5, 10 and 20.}
\end{center}
\end{figure}

\end{document}